\documentclass[twocolumn,aps,showpacs,amsmath,amssymb, nofootinbib]{revtex4}
\usepackage{amssymb}
\usepackage{graphicx}
\usepackage{dcolumn}
\usepackage{mathrsfs}
\usepackage{bm}
\usepackage{lipsum}
\usepackage{amsmath}

\DeclareMathOperator{\Arth}{Arth}

\begin{document}
\title{Role of Brownian motion and N\'{e}el relaxations \\in M\"{o}ssbauer spectra of magnetic liquids   }

\author{A.Ya.Dzyublik}
\email[E-mail: ]{dzyublik@ukr.net}

\author{ V.Yu.Spivak}
\affiliation{
Institute for Nuclear Research, National Academy of
Sciences of Ukraine, avenue Nauki 47,
03680  Kyiv, Ukraine}
\date{\today}
\begin{abstract}
The absorption cross section of M\"{o}ssbauer radiation in magnetic liquids is calculated, taking into consideration both translational and rotational Brownian motion of magnetic nanoparticles. Stochastic reversals of their magnetization are also regarded in the absence of external magnetic field. The role of Brownian motion in ferrofluids is considered  in the framework of the diffusion theory, while for the magnetorheological fluids with large nanoparticles it is done in the framework of the Langevin's approach. For rotation we derived the equation analogous to Langevin's equation and gave the corresponding correlation function. In both cases the equations for rotation are solved in the approximation of small rotations during lifetime of the excited state of M\"{o}ssbauer nuclei. The influence  of magnetization relaxations is studied with the aid of the Blume-Tjon model.
   \end{abstract}

\pacs{06.20.-f, 42.55.-f, 32.90.+a, 03.5.Nk}

\maketitle
\section{Introduction}
Suspensions of magnetic nanoparticles (MNPs)  attract great attention due to their numerous applications in  technique, medicine and biology [1-17]. It is provided by large magnetic moment of MNPs, which allows to manipulate them by moderate magnetic fields. Depending on the dimensions of MNPs, they can be divided into   magnetorheological fluids  formed by MNPs with the diameter of the order of $1 \mu$m and ferrofluids  with
 dimensions of  MNPs $\sim 10$ nm (see, e.g., \cite{Brazil}).
Viscosity  of the magnetorheological liquids, being subjected to the magnetic field,  enormously increases,  so that  they may even transform into a solid body.  This property gives possibility to  use such suspensions  in dampers, brakes and clutches \cite{Wang}.

Ferrofluids are widely used in computers,
loudspeakers, semiconductors, motion controllers, sensors, ink-jet printers,  seals, bearings, stepper motors, etc [1-8].

In medicine ferrofluids are  employed in hyperthermia \cite{Mody}, for drug delivering to local ill regions of the body
 and as  contrast agents  in magnetic resonance imaging (MRI)  \cite{med}.
Recently, much more progressive method of magnetic particle
imaging (MPI) was developed for visualizing
MNPs in humans and animals [11-15].
The advantage of MPI is that it is more fast, quantitative, and sensitive than MRI.

Note that MNPs are always  coated with a polymer shell to prevent their  agglomeration in a solution.
The commercial ferrofluids are predominantly based on magnetite Fe$_3$0$_4$ particles.

Usually MNPs have single easy-magnetization axis $\zeta$ and their magnetization ${\bf M}$ tends to be oriented along it or in the opposite direction, keeping the constant value $|{\bf M}|$. The anisotropy potential energy of such  particles in the absence of external magnetic fields, versus the angle $\Theta$ between ${\bf M}$ and  axis $\zeta$,  is  represented by two potential wells at $\Theta=0$ and $\pi$, separated by the potential barrier:
\begin{equation}\label{eq:pot}
  E=K_{\textrm{\scriptsize{eff}}}V\sin^2\Theta,
\end{equation}
where  $K_{\textrm{\scriptsize{eff}}}$ denotes  the effective magnetic anisotropy, $V$ the particle volume. The magnetization, oscillating in one of the potential wells, from time to time gets sufficient energy to jump over the barrier into the neighboring well.
In the symmetric potential (\ref{eq:pot})  the magnetization reversals occur with equal rate   $w=1/\tau$ in both sides, where the relaxation time $\tau_N$
is determined by N\'{e}el's formula \cite{Neel}
\begin{equation}\label{}
\tau_N=\tau_0\exp{\left( \frac{K_{\textrm{\scriptsize{eff}}}V}{kT}\right)}
\end{equation}
with the constant factor $\tau_0 \sim 10^{-9} - 10^{-13} s^{-1}$ (see, e.g., \cite{Landers}).

Effectiveness of the MNPs in different applications strongly depends on such parameters as the N\`{e}el relaxation time and Brownian rotational relaxation time of MNPs  as well as their relation to temperature and viscosity of the carrier fluid. The M\"{o}ssbauer spectroscopy is the most powerful method allowing us to determine all such characteristics.
Soon after discovery the M\"{o}ssbauer effect has been  applied for investigation of the Brownian motion of nanoparticles in liquids [18-26]. Foundation of these studies has been laid by Singwi and Sj\"{o}lander \cite{Singwi}, who expressed the absorption cross section of  M\"{o}ssbauer  rays by chaotically moving nanoparticle in terms of the Van Hove auto-correlation function. Having described the translational motion of the Brownian particle by  diffusion equation, they  found that the broadening of  M\"{o}ssbauer line linearly depends on the ratio of temperature $T$ and viscosity of the liquid $\eta$. This conclusion was supported experimentally for small nanoparticles [18-24].  But somewhat later for large particles it was observed considerable curvature of the line, which was explained theoretically in Refs.~\cite{K,B2}, where the Brownian motion was described by means of the stochastic  Langevin's equation.

For the first time the role of the Brownian rotation was analyzed by Zatovski\v{i} \cite{Zat},  who found most strict  solution for the absorption cross section.
In Ref.~\cite{Dz1} this  task has been solved in small-angle approximation, taking into account that during the lifetime of the excited  M\"{o}ssbauer nucleus the root mean-square angle of the Brownian rotation is usually much less than unity. Such simple calculations were expanded to the case of ellipsoidal Brownian particles in Ref.~\cite{Dz2}. Another more cumbersome approach to the problem was developed in Ref.~\cite{Af}.

Besides, manifestation of the N\'{e}el relaxations along with the Brownian motion in M\"{o}ssbauer spectra was studied in Ref.~\cite{Dz3}, whereas
Landers {\it et al.} [17] seem to be the first who observed the  M\"{o}ssbauer spectra in ferrofluids. Interesting experimental results have been also reported in Refs.[33-36].

In this paper we research in much more detail the impact of Brownian rotation together with translational diffusion on the shape of  M\"{o}ssbauer spectra.
In particular, for the first time the Brownian rotation of large nanoparticles will be regarded in the framework of  Langevin's formalism on the same footing as the translational motion.

For ferrofluids we shall apply a simplest relaxation model, when the magnetization vector of the particle ${\bf M}(t)$ makes stochastic jumps between the values ${\bf M}$ and $-{\bf M}$ along the easy axis $\zeta$.   Respectively, the magnetic field at the nucleus, being antiparallel to ${\bf M}(t)$, takes the values ${\bf h}_0(t)={\bf h}_0f(t)$ with $f(t)=\pm 1$. The magnetic field ${\bf h}_0$ causes splitting of the nuclear sublevels giving rise to a Zeeman sextet. For generality we adopt that along the field ${\bf h}_0$ there is an electric field gradient,  which ensures a quadrupole splitting of the lines.
This model was previously applied to calculations of M\"{o}ssbauer spectra by Blume and Tjon \cite{Blume}.

\section{Basic equations}
In order to separate the translational and rotational motion we first introduce the coordinate frame $\{x,\;y,\;z\}$ with the origin in the center of the particle and axis $z$ along the beam of incident $\gamma$-quanta. In addition, we introduce the frame $\{\xi,\;\eta,\;\zeta \}$ with  axis $\zeta$ along the easy-magnetization  axis of the particle.
Position of the M\"{o}ssbauer nucleus $^{57}$Fe in the laboratory frame $\{x_L,\; y_L,\;z_L\}$ is determined by the radius-vector
\begin{equation}\label{eq:r1}
{\bf X}={\bf R}+{\bf r}+{\bf u},
\end{equation}
where the vector ${\bf R}$ indicates position of the center of the Brownian particle, ${\bf r}$ specifies the equilibrium position of the nucleus
in the frame $\{x,\;y,\;z\}$, and ${\bf u}$ is the  displacement from this cite.

Random reversals of the magnetization ${\bf M}$ and the Brownian motion are independent processes. Therefore the absorption cross section  of $\gamma$-quanta with the energy  $E=\hbar\omega$ and wave vector ${\boldsymbol \kappa}$ by the M\"{o}ssbauer nucleus
$^{57}$Fe, embedded in the Brownian particle, may be written as \cite{Dz3}
\begin{eqnarray}\label{eq:s}
\sigma_a(\omega)= \frac{\sigma_0\Gamma_a}{2}e^{-2W} \\ \times
\mbox{Re}\int_0^{\infty}\frac{dt}{\hbar}e^{i(\omega-\omega_a)t-\Gamma_at/2\hbar}
G_B({\boldsymbol \kappa},t)G_{N}(t), \nonumber
\end{eqnarray}
where $\sigma_0$ is the resonant value of the absorption cross section of $\gamma$-quanta by a fixed nucleus in the absence of the hyperfine structure, $E_a=\hbar\omega_a$ and $\Gamma_a$ are the energy and width of the resonant level of the absorbing nucleus,
$e^{-2W}$ is the Debye-Waller factor, $G_B({\boldsymbol \kappa},t)$  denotes the  Fourier-transform of the classical autocorrelation function for the Brownian motion,
$G_{N}(t)$ the correlation function for the  N\'{e}el relaxations of magnetization.

This cross section is to be averaged over the energy  distribution of $\gamma$-quanta emitted by a source without recoil
\begin{equation}\label{}
w_e(E)=\frac{2\pi}{\Gamma_e}\frac{1}{(E-E_e-s)^2+(\Gamma_e/2)^2},
\end{equation}
where $s=vE_e/c$ denotes the Doppler shift for a source, moving with the velocity $v$ relative to an absorber.
Then experimentally measured cross section takes the form
\begin{eqnarray}\label{eq:s3}
\sigma_a(s)= \frac{\sigma_0\Gamma_a}{2}e^{-2W} \\ \times
\mbox{Re}\int_0^{\infty}\frac{dt}{\hbar}e^{ist/\hbar-\Gamma_0 t/2\hbar}
G_B({\boldsymbol \kappa},t)G_{N}(t), \nonumber
\end{eqnarray}
where $\Gamma_0=\Gamma_e+\Gamma_a$ means the width observed when any broadening due to Brownian motion or N\'{e}el  relaxations is absent.

For spherical particles the translational and rotational Brownian motions are separated, so that
\begin{equation}\label{}
G_B({\boldsymbol \kappa},t)=G_t({\boldsymbol \kappa},t)G_r({\boldsymbol \kappa},t),
\end{equation}
where $G_t({\boldsymbol \kappa},t)$ and $G_r({\boldsymbol \kappa},t)$ are the Fourier transforms of the correlation functions for translational motion and rotation, respectively.

\section{Correlation functions}
In this section we shall give the correlation functions for the translational and rotational Brownian  motion of spherical nanoparticles in a liquid, provided by corresponding diffusion equations. Besides, the correlator responsible for the N\'{e}el relaxations of the MNPs  magnetization, derived in Ref.~\cite{Blume}, will be reproduced below in somewhat changed form.

\subsection*{3.1 Translational Brownian motion}
For the translational Brownian motion, described by simple diffusion equation, the  correlation function has the form  \cite{Singwi}
\begin{equation}\label{eq:Cor}
G^t_s({\bf R},t)=(4\pi D_t|t|)^{-3/2}\exp[- R^2/4D_t|t|].
\end{equation}
Here we suppose that at the initial moment $t=0$ the  particle is located in the origin of the laboratory frame.
The Fourier-transform of the function (\ref{eq:Cor}) reads
\begin{equation}\label{eq:G}
G_s^t({\boldsymbol \kappa},t)=\exp(-\kappa^2D_t|t|).
\end{equation}
As a result,  the broadening of the M\"{o}ssbauer line caused by translational diffusion of a spherical nanoparticle is given by \cite{Singwi}
\begin{equation}\label{}
\Delta\Gamma_t=2\hbar\kappa^2D_t,
\end{equation}
where the  translational diffusion coefficient
\begin{equation}\label{}
D_t=\frac{kT}{6\pi \eta a_h},
\end{equation}
 $\eta$ is the viscosity coefficient of the liquid, $a_h$ the  hydrodynamic radius of the nanoparticle being a sum of the core radius $a_c$ and a thickness of its coating $d$.

\subsection*{3.2 Brownian rotation}
The mean-square angle of rotation $\Delta_r^2$ of the Brownian particle in a liquid during time $t$ is determined by Einstein's formula  \cite{Ein}
\begin{equation}\label{eq:Ein}
\Delta^2_r = 2D_rt,
\end{equation}
depending on the rotation diffusion coefficient
\begin{equation}\label{eq:Ein2}
D_r=\frac{kT}{8\pi\eta a_h^3},
\end{equation}
where $a_h$ is the hydrodynamic radius of the particle.

Let us estimate the  $\Delta^2_r$ for rotation during the time $t$ of the order of the lifetime $\tau_N=\hbar/\Gamma_a=141$~ns for $^{57}$Fe. We take  the  parameters of the experiment \cite{Landers}, which correspond to maximal value of  $\Delta^2_r$: $a_h=7$ nm and $\eta=22.5$~cp (viscosity of the 70\% glycerol solution at $T=293$~K). In this case $\Delta^2_r\approx 4\cdot 10^{-3}$. In all other measurements \cite{Landers}, corresponding to lower temperatures and larger particles,  $\Delta^2_r$ is much less. Thus, we can really treat the Brownian rotation in the small-angle approximation.

The Fourier-transform of the rotational correlation function is calculated with the aid of the probability density   $W({\bf n}, {\bf n}_0;t)$
 of the  Brownian  rotation from ${\bf n}_0$ to ${\bf n}$ during time $t$:
\begin{eqnarray}\label{eq:G12}
G_s^r({\boldsymbol \kappa},t)=\int d{\bf n}W({\bf n}, {\bf n}_0;t) e^{i{\boldsymbol \kappa}({\bf n}-{\bf n}_0)r},
\end{eqnarray}
where orientation of  the unit vectors ${\bf n}_0={\bf r}(0)/r$ and ${\bf n}={\bf r}(t)/r$ in the frame $\{x,\;y,\;z\}$ are determined by the spherical angles  $\theta_0, \varphi_0$ and $\theta, \varphi$, respectively.
 The function $W({\bf n}, {\bf n}_0;t)$ is looked for as a solution of the rotational diffusion equation \cite{Leon}
\begin{equation}\label{eq:W1}
\frac{\partial W}{\partial t}=D_r\left[\frac{\partial^2 W}{\partial \theta^2}+\cot\theta\frac{\partial W}{\partial \theta}
+\frac{1}{\sin^2\theta}\frac{\partial^2 W}{\partial\varphi^2}  \right]
\end{equation}
with the initial condition
\begin{equation}\label{eq:W2}
W({\bf n}, {\bf n}_0;0)=\delta({\bf n}-{\bf n}_0).
\end{equation}
The probability of all possible events equals unity, therefore the probability density is normalized as
\begin{equation}\label{}
\int_0^{\pi}\sin\theta d\theta\int_0^{2\pi}d\varphi
W({\bf n}, {\bf n}_0;t)=1.
\end{equation}

Let us introduce one more frame $x',\;y',\;z'$, whose axis $z'$ is directed along ${\bf n}_0$. The orientation of the vector ${\bf n}$ in this frame is determined by the spherical angles $\vartheta, \phi$. In the approximation of small Brownian rotations, when
\begin{equation}\label{eq:50}
({\bf n}-{\bf n}_0)^2\approx \vartheta^2<<1,
\end{equation}
the basic equation~(\ref{eq:W1}) transforms to
\begin{equation}\label{eq:W4}
\frac{\partial W}{\partial t}=D_r\left[\frac{\partial^2 W}{\partial \vartheta^2}+\frac{1}{\vartheta}\frac{\partial W}{\partial \vartheta}
+\frac{1}{\vartheta^2}\frac{\partial^2 W}{\partial\phi^2}  \right].
\end{equation}
Notice that the same form has the equation, which describes  diffusion of the point-like particle on the plane, expressed  in polar coordinates with the radial coordinate $\vartheta$ and azimute angle $\phi$. Hence, a solution of Eq.~(\ref{eq:W4}), satisfying the initial condition (\ref{eq:W2}),   is
\begin{equation}\label{eq:W5}
 W({\bf n}, {\bf n}_0;t)= \frac{1}{4\pi D_r |t|}\exp\left(-\frac{\vartheta^2}{4D_r |t|} \right).
\end{equation}
From here we see that the mean-square rotation angle is really determined by Eq.~(\ref{eq:Ein}).

In order to calculate the Fourier-transform of $ W({\bf n}, {\bf n}_0;t)$ we shall express the components  $n_{x},\;n_y,\;n_{z}$ of the of unit vector ${\bf n}$ in spherical angles:
\begin{equation}\label{eq:51}
n_x=\sin\theta\cos\varphi, \quad n_y=\sin\theta\sin\varphi, \quad n_z=\cos\theta.
\end{equation}
Simple calculation gives
\begin{equation}\label{eq:52}
({\bf n}-{\bf n}_0)^2 = (\theta-\theta_0)^2+\sin^2\theta_0(\varphi-\varphi_0)^2.
\end{equation}
Comparing (\ref{eq:50}) with (\ref{eq:52}) we rewrite (\ref{eq:W5}) as
\begin{eqnarray}\label{eq:53}
 W({\bf n}, {\bf n}_0;t)= \frac{1}{4\pi D_r |t|} \qquad\qquad \\ \times
 \exp\left(-\frac{(\theta-\theta_0)^2}{4D_r |t|} \right)\exp\left(-\frac{\sin^2\theta_0(\varphi-\varphi_0)^2}{4D_r |t|} \right).\nonumber
\end{eqnarray}

Then starting from the formula
\begin{eqnarray}\label{eq:54}
G_s^r({\boldsymbol \kappa},t)=
\int_{-\infty}^{\infty}\sin\theta d\theta\exp[-i\kappa r\sin\theta_0(\theta-\theta_0)]\nonumber \\
\times \int_{-\infty}^{\infty}d\varphi W({\bf n},{\bf n}_0;t),
\end{eqnarray}
we arrive at the Fourier-transform of the rotation correlation function:
\begin{equation}\label{eq:Gr}
G_s^r({\boldsymbol \kappa},t)= \exp\left[-\kappa^2D_r r^2\sin^2\theta_0 |t|  \right].
\end{equation}

\subsection*{3.3 Magnetization relaxations}
Following Ref.~\cite{Blume} we suppose that there is an electric field gradient along the magnetic field  ${\bf h}_0$ at the nucleus $^{57}$Fe.
The constant field ${\bf h}_0$ gives rise to Zeeman splitting of sublevels  $1/2, M_g$ and  $3/2,M_e$  of the ground and excited nuclear states, respectively.
Here $M_g$ and $M_e$ are the projections of the nuclear spin on the direction ${\bf h}_0$.
In the fluctuating field ${\bf h}_0(t)$ the M\"{o}ssbauer spectrum is described by the correlator $G_{N}(t)$  \cite{Blume}
\begin{eqnarray}\label{eq:F1}
G_{N}(t)=\sum_{j=1}^6J_{j}(\beta)\exp[-iQ(3M_e^2-15/4)t/\hbar]\\
\times \left(\exp\left[i\alpha_{j}\int_0^{t}f(t')dt' \right]\right)_{av},\nonumber
\end{eqnarray}
where  $(...)_{av}$ implies the stochastic averaging, $f(t)=\pm 1$, the parameter $Q$ determines a quadrupole shift of the lines, the factors $J_{j=M_g \to M_e}(\beta)$ are  relative intensities of the Zeeman sextet:
\begin{eqnarray}
J_{1=-1/2 \to -3/2}(\beta)=J_{6=1/2 \to 3/2}(\beta)=\frac{3}{16}(1+\cos^2\beta),\nonumber \\
J_{2=-1/2 \to -1/2}(\beta)=J_{5=1/2 \to 1/2}(\beta)=\frac{1}{4}\sin^2\beta,\qquad\\
J_{3=-1/2 \to 1/2}(\beta)=J_{4=1/2 \to -1/2}(\beta)=\frac{1}{16}(1+\cos^2\beta),\nonumber
\end{eqnarray}
 depending on the angle $\beta$ between the wave vector of $\gamma$-quanta ${\boldsymbol \kappa}$ and magnetization $+{\bf M}$.
Here the lines of the Zeeman sextet are numerated in the order of  growing energy. In the absence of external magnetic fields,
when the particles are oriented randomly, the averaged relative intensities of the lines
$J_i=\langle J_i(\beta)\rangle$ are
\begin{equation}\label{}
J_1=J_6=\frac{3}{12},\;\; J_2=J_5=\frac{2}{12},\;\;J_3=J_4=\frac{1}{12}.
\end{equation}

 Then the stochastic averaging results in \cite{Blume}
\begin{eqnarray}\label{eq:R1}
\left(\exp
\left[i\alpha_j\int_0^{t}f(t')dt' \right]\right)_{av}=\\
= (\cos x_jwt+x_j^{-1}\sin x_jwt)\exp(-wt),\nonumber
\end{eqnarray}
with parameters
\begin{eqnarray}\label{}
x_j=[(\alpha_{j}/ w)^2-1]^{1/2},\nonumber \\
 \alpha_j=\alpha_{eg}=(g_gM_g-g_eM_e)\mu_Nh_0/\hbar,
\end{eqnarray}
depending on the nuclear magneton $\mu_N$ and gyromagnetic ratios  $g_g$,  $g_e$ of the ground and excited states, respectively.
From now on, for brevity, we shall omit the exponential in Eq.~(\ref{eq:R1}),
 associated with the quadrupole splitting. Once $Q\neq  0$, in all equations below the Doppler shift $s$ is to be replaced by $s-Q(3M_e^2-15/4)$.

\section{Absorption cross section}
Substituting (\ref{eq:G}), (\ref{eq:Gr})  and (\ref{eq:R1}) into (\ref{eq:s3})
one finds the absorption cross section
\begin{widetext}
\begin{equation}\label{eq:s1}
\sigma_a(s)=\frac{\sigma_0\Gamma_a }{4}e^{-2W}\mbox{Re}\sum_{j=1}^6  J_j \left[\left(1-\frac{i}{x_j}\right)
\frac{i}{s+x_j\hbar w+i\Gamma_{\textrm{\scriptsize{eff}}} /2}
+ \left(1+\frac{i}{x_j}\right)\frac{i}{s-x_j\hbar w+i\Gamma_{\textrm{\scriptsize{eff}}} /2}\right],
\end{equation}
\end{widetext}
where the effective width
\begin{equation}\label{eq:G5}
\Gamma_{\textrm{\scriptsize{eff}}} =\Gamma_0+\Delta\Gamma_B+\Delta\Gamma_N,
\end{equation}
 $\Delta\Gamma_N= 2\hbar w$  is a broadening due to the N\'{e}el relaxations,
while the Brownian broadening is given by a sum
\begin{equation}
 \Delta\Gamma_B=\Delta\Gamma_t+\Delta\Gamma_r({\bf r})
 \end{equation}
  with the rotational contribution
 \begin{equation}\label{}
\Delta\Gamma_r({\bf r})= 2\hbar\kappa^2D_r r^2\sin^2\theta_0,
\end{equation}
depending on the coordinates of the M\"{o}ssbauer nucleus $^{57}$Fe.

For uniform distribution of these nuclei   in nanoparticle the averaged cross section is defined by
\begin{equation}\label{eq:av}
\langle \sigma_a(s)\rangle = \frac{3}{2a_c^3}\int_0^{a_c}r^2 dr\int_0^{\pi}\sigma_a(s)\sin\theta_0d\theta_0.
\end{equation}
Having substituted here the expression (\ref{eq:s1}) we introduce new variables $\xi=\cos\theta_0$ and $\rho=r/a_c$ to obtain
\begin{eqnarray}
\langle\sigma_a(s)\rangle=\frac{3\sigma_0\Gamma_a }{4}e^{-2W}\mbox{Re}\sum_{j=1}^6  J_j \int_0^1\rho^2 d\rho \\ \times
\left[\left(1-\frac{i}{x_j}\right)I_j^{+}(\rho)+ \left(1+\frac{i}{x_j}\right)I_j^{-}(\rho) \right],\nonumber
\end{eqnarray}
where $I_j^{\pm}(\rho)$ stands for the integral
\begin{equation}\label{}
I_j^{\pm}(\rho)=\int_{0}^1\frac{d\xi}{A_j^{\pm}(\rho)-B(\rho)\xi^2}
\end{equation}
with
\begin{eqnarray}\label{eq:B}
B =i\hbar\kappa^2D_ra_c^2\rho^2= (3i/8)\Delta\Gamma_t(a_c/a_h)^2\rho^2
\end{eqnarray}
and
\begin{equation}\label{eq:A}
A_j^{\pm}= s \pm x_jw +i(\Gamma_0+\Delta\Gamma_t+\Delta\Gamma_N)/2+B.
\end{equation}
Trivial integration gives
\begin{equation}\label{}
I_j^{\pm}(\rho)= \frac{1}{\sqrt{A_j^{\pm}B}}\Arth\left({\sqrt{B/A_j^{\pm}}}\right).
\end{equation}

In the case of slow relaxations, when $ w<<|\alpha_{j}|$ and respectively $x_j>>1$ as well as $x_j w\approx \alpha_{j}$, the cross section reduces to Zeeman's pattern with broadened lines:
\begin{eqnarray}
\langle\sigma_a(s)\rangle=\frac{\sigma_0\Gamma_a}{2}e^{-2W_a}\mbox{Re} \sum_{j=1}^6 J_{j} \\
\times \int_0^1\rho^2d\rho\frac{1}{\sqrt{A_jB}}
\Arth\left(\sqrt{B/A_j}\right), \nonumber
\end{eqnarray}
where $B$ is again determined  by formula (\ref{eq:B}), while  $A_j$ takes the form
\begin{equation}\label{eq:2A}
A_j=s-\hbar\alpha_j +i(\Gamma_0+\Delta\Gamma_t+\Delta\Gamma_N)/2+B.
\end{equation}

In the opposite limit of very rapid relaxations as $w>>|\alpha_{j}|$ the nucleus only feels an average zero magnetic field. In this case the cross section collapses to single line \footnote{if $Q\neq 0$ the spectrum collapses  to a quadrupole doublet.}
\begin{eqnarray}\label{eq:10}
\langle\sigma_a(s)\rangle=\frac{3\sigma_0\Gamma_a }{2}e^{-2W_a}\mbox{Re}\int_0^1\rho^2 d\rho \\
\times\frac{1}{\sqrt{AB}}\Arth\left(\sqrt{B/A}\right), \nonumber
\end{eqnarray}
where $B$ remains the same, while  $A$ becomes
\begin{equation}\label{eq:2A}
A=s +i(\Gamma_0+\Delta\Gamma_t)/2+B.
\end{equation}
Note also that the same expression (\ref{eq:10}) describes the M\"{o}ssbauer spectra of nonmagnetic Brownian particles.

The formulas considerably simplify, if we average  only  $\Delta\Gamma_r({\bf r})$ instead of the  whole cross section (\ref{eq:s1}).
Then
\begin{equation}\label{eq:21}
\langle r^2\sin^2\theta_0\rangle = 0.4a_c^2
\end{equation}
and
\begin{equation}\label{}
\langle\Delta\Gamma_r({\bf r})\rangle=0.6\hbar\kappa^2D_t(a_c/a_h)^2.
\end{equation}
  In this case the averaged cross section is determined by the same formula (\ref{eq:s1}) but with the Brownian broadening $\Delta\Gamma_B$ replaced by
\begin{equation}\label{eq:22}
\langle\Delta\Gamma_B\rangle = 2\hbar\kappa^2D_t[1+0.3(a_c/a_h)^2].
\end{equation}
  As to the  cross section (\ref{eq:10}, it reduces to
\begin{equation}\label{eq:SS2}
 \langle\sigma_a(s)\rangle=\frac{\sigma_0\Gamma_a }{4}e^{-2W_a}
 \frac{\Gamma_0+\langle \Delta\Gamma_B\rangle}{s^2+(\Gamma_0+\langle \Delta\Gamma_B\rangle)^2/4}.
\end{equation}
 If a  contribution of the rotation is neglected, Eq.~(\ref{eq:SS2}) coincides with the result of Singwi and Sjolander \cite{Singwi}.

The role of  rotational diffusion is illustrated by Fig.1, where all the cross sections  are calculated in units $(3\sigma_0\Gamma_a /2)e^{-2W}$ as a function of the dimensionless parameter $2s/(\Gamma_0+\Delta\Gamma_{t})$ for the case, when  $\Delta\Gamma_{t}=4\Gamma_0$ and $a_c=a_h$.
The cross section (\ref{eq:10}) is drawn by the solid line.  The Singwi-Sjolander's curve, given by Eq.~(\ref{eq:SS2}) with rotational contribution $\langle\Delta\Gamma_r({\bf r})\rangle=0$,  by the dashed one. In addition,  the approximate curve (\ref{eq:SS2}) with $\langle\Delta\Gamma_B\rangle=2.6\hbar\kappa^2D_t$, is represented by the dash-dotted line. It is seen that it  surprisingly well approximates the exact result (\ref{eq:10}).
\begin{figure}[t]
\label{fig1}\vspace{-1cm}
\centerline{\includegraphics[height= 8cm, width= 9cm]{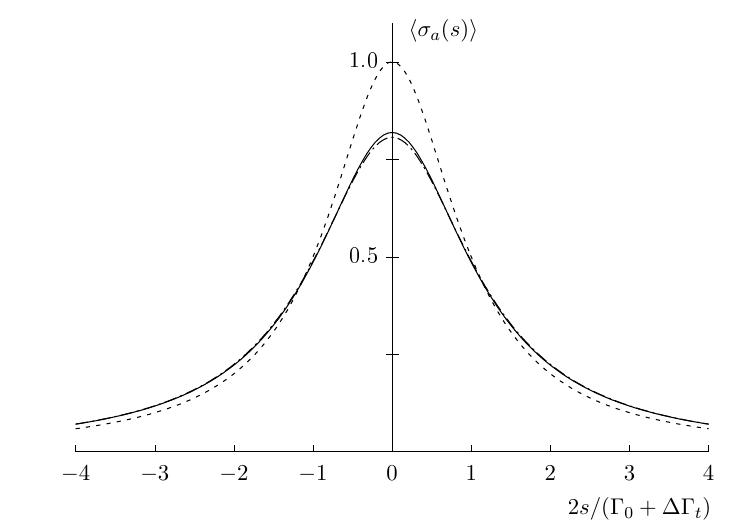}}
\vspace{-0.4cm}
\caption{Dependance  of the  absorption cross sections on the Doppler shift $s$, expressed in units $(\Gamma_0+\Delta\Gamma_t)/2$. The cross section calculated by Eq.~(\ref{eq:10}) is drawn by solid line,  the result of Singwi and Sjolander \cite{Singwi} by dashed one, the approximate expression (\ref{eq:SS2}) by dash-dotted. }
\end{figure}

\section{Approach based on Langevin's equation}
The correlation function (\ref{eq:Cor}) is not valid at small times. More correctly self-diffusion is described by  the correlation function \cite{Singwi, Chan}
\begin{equation}\label{eq:G2}
G_t({\bf R}, t)=[2\pi\rho_t(t)]^{-3/2}\exp\left[-\frac{R^2}{2\rho_t(t)}\right],
\end{equation}
where
\begin{equation}\label{eq:rho}
\rho_t(t)=\frac{2D_t}{\beta_t}[\beta_t t-1+\exp(-\beta_t t)],
\end{equation}
with
\begin{equation}\label{}
\beta_t=6\pi a \eta/m=\frac{kT}{D_tm}.
\end{equation}
The parameter $\beta_t^{-1}$ means the characteristic (relaxation) time   for the Brownian translational motion.

At $t>> \beta_t^{-1}$ the correlation functions (\ref{eq:Cor}) and (\ref{eq:G2}) coincide. In the opposite limit of $t<< \beta_t^{-1}$ an employment of
diffusion approach leads to paradox, remarked in Ref.~\cite{Chuev1}.  Really, from the diffusion equation it follows that at $t\geq 0$ the mean-square displacement
$\langle x^2 \rangle=2D_tt$, and therefore the root mean-square velocity along the axis $x$, given by ${\sqrt{\langle v^2_x \rangle}}= \sqrt{2D_t/t}$, tends to infinity if $t\to 0$.
At the same time, in correspondence with (\ref{eq:rho}), at $t<<\beta_t^{-1}$ the mean-square displacement $\langle x^2 \rangle = D_t\beta_t t^2$, hence
$\sqrt{\langle v^2_x \rangle}= \sqrt{D_t\beta_t}=(kT/m)^{1/2}$. Thus, the mean kinetic energy of the Brownian particle $\bar{E}_{\textrm{\scriptsize{kin}}}$ at $t \to 0$ occurs to be determined by the same expression as $\bar{E}_{\textrm{\scriptsize{kin}}}$ for the molecules of the ideal gas:
\begin{equation}\label{}
\bar{E}_{\textrm{\scriptsize{kin}}}=3kT/2.
\end{equation}

The function (\ref{eq:G2}) was derived by Chandrasekhar \cite{Chan} from Langevin's equation
\begin{equation}\label{eq:L}
\frac{d^2{\bf R }}{dt^2}=
-\beta_t\frac{d{\bf R}}{dt}+{\bf F}_{\textrm{\scriptsize{rand}}}(t)/m.
\end{equation}
Here on the right-hand side  the first term is responsible for the dynamical friction,
${\bf F}_{\textrm{\scriptsize{rand}}}(t)/m$ are random forces acting on the Brownian particle.

Let us find now analogous correlation function for the rotational Brownian motion. For this aim we shall first derive the equation similar to Eq. ({\ref{eq:L}),
starting from well-known relationship for the angular momentum ${\bf L}$ of the rotating rigid body and the total  torque ${\bf K}$ acting on it \cite{Landau}:
\begin{equation}\label{eq:La}
\frac{d{\bf L}}{dt}={\bf K}.
\end{equation}
 We take into account that the angular momentum for a rigid sphere of the radius $a$ is bound to its angular frequency of rotation  ${\boldsymbol\omega}$ by
 \begin{equation}\label{eq:L1}
 {\bf L}=  \mathfrak {J}{ \boldsymbol \omega},
 \end{equation}
 where the inertia moment of the  sphere
  \begin{equation}\label{}
  \mathfrak {J}=0.4ma_h^2.
 \end{equation}

The ${\bf K}$ equals  a sum of the friction torque  \cite{Kirch}
\begin{equation}\label{eq:L2}
{\bf K}_{\textrm{\scriptsize{fr}}}=-8\pi\eta a_h^3 {\boldsymbol\omega}.
\end{equation}
 and  torques  ${\bf K}_{\textrm{\scriptsize{rand}}}(t)$ due to random forces.

Inserting (\ref{eq:L1}), (\ref{eq:L2}) into (\ref{eq:La}) one gets the equation, governing the stochastic rotational motion:
\begin{equation}\label{eq:D}
\frac{d{\boldsymbol \omega}}{dt}
=-\beta_r{\boldsymbol \omega} +{\bf K}_{\textrm{\scriptsize{rand}}}(t)/\mathfrak {J},
\end{equation}
where
\begin{equation}\label{}
\beta_r=\frac{20\pi\eta a_h}{m}=\frac{5}{2}\frac{kT}{D_r ma_h^2 }.
\end{equation}

Keeping in mind that for  small rotations ${\boldsymbol \omega}$ is perpendicular to ${\bf n}_0$ and equals $\omega=d\vartheta/dt$, we
transform (\ref{eq:D}) to the equation
\begin{equation}\label{eq:D2}
\frac{d^2\vartheta}{dt^2}
=-\beta_r\frac{d\vartheta}{dt} +  K_{\textrm{\scriptsize{rand}}}(t)/\mathfrak {J},
\end{equation}
formally  equivalent to Langevin's equation (\ref{eq:L}) in  one-dimensional case. Here in the same approximation we  ignore the boundary conditions for
the angle $\vartheta$ and accept that it ranges from
 $-\infty$ to $\infty$. Further  repeating derivation, done by Chandrasekhar \cite{Chan}, one gets the correlation function
\begin{equation}\label{eq:Gt}
G_r(\vartheta, t)=[2\pi\rho_r(t)]^{-1/2}\exp[-\vartheta^2/2\rho_r(t)],
\end{equation}
where the function $\rho_r(t)$ is again defined by Eq.~(\ref{eq:rho}), but with index $t$ replaced by $r$. As to the Fourier-transform, it is given now by
\begin{equation}\label{}
G_r({\boldsymbol \kappa},t)=\exp\left[-\kappa^2\rho_r(t) r^2\sin^2\theta_0/2  \right].
\end{equation}

  Combining these equations we get in the slow-relaxation limit the cross section as a superposition of six lines:
\begin{equation}
\sigma_a(s)=\sum_{j=1}^6\sigma_a^{(j)}(s),
\end{equation}
each of them is given by
\begin{eqnarray}\label{eq:111}
\sigma_a^{(j)}(s)=\frac{\sigma_0\Gamma_a}{4}e^{-2W}e^{b_t+b_r} J_{j} \\ \times
\sum_{n,k=0}^{\infty}\frac{(-b_t)^n}{n!}\frac{(-b_r)^k}{k!}
\frac{\Gamma_{n,k}}{(s-\hbar\alpha_{j})^2   +\left(\frac{\Gamma_{n,k}}{2}\right)^2}, \nonumber
\end{eqnarray}
where we introduced the parameters
\begin{equation}\label{}
b_t=\frac{\kappa^2D_t}{\beta_t}, \qquad b_r=\frac{\kappa^2D_rr^2\sin^2\theta_0}{\beta_r}
\end{equation}
and  the widths
\begin{equation}\label{}
\Gamma_{n,k} =\Gamma_{\textrm{\scriptsize{eff}}} +2n\hbar\beta_t+2k\hbar\beta_r.
\end{equation}

For estimations it is sufficient to replace averaging of the cross section  over ${\bf r}$ by the averaging of
$r^2\sin^2\theta_0$  i.e., to replace this product by
$\langle r^2\sin^2\theta_0\rangle = 0.4a_c^2$. Then one has the relation
\begin{equation}\label{}
\langle b_r({\bf r}) \rangle =0.045 b_t,
\end{equation}
which allows to   set in (\ref{eq:111}) $b_r=0$.

\begin{figure}[ht]
\label{fig2}\vspace{-1cm}
\centerline{\includegraphics[height= 6cm, width= 9cm]{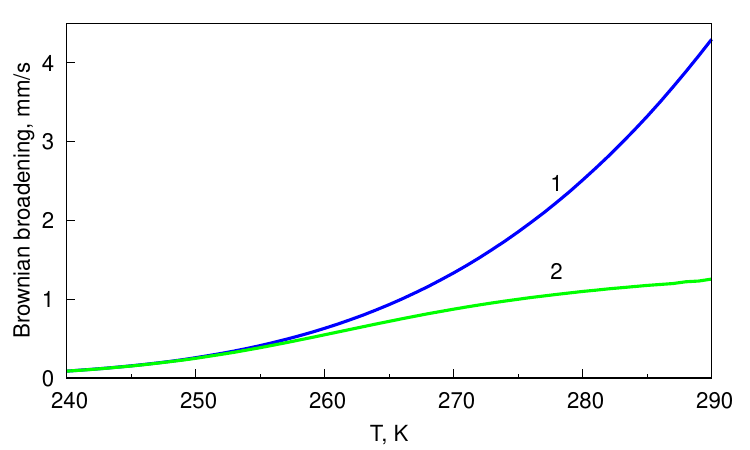}}
\vspace{-0.3cm}
\caption{ Brownian broadening of the absorption  lines vs temperature for the magnetite nanoparticles with diameter $a_h=a_c=700$ nm in the 60\% glycerol-water mixture, calculated in the diffusion approach (1) and  Langevin's one (2)  }
\end{figure}

In order to find now the integral width of $j$th Zeeman line we employ standard formula
\begin{equation}\label{}
\Gamma_{\textrm{\scriptsize{int}}} =\frac{2}{\pi \sigma^{(j)}_a(\hbar\alpha_{j})}\int_{-\infty}^{\infty}\sigma^{(j)}_a(s)ds.
\end{equation}
Simple calculation gives
\begin{eqnarray}\label{eq:101}
\Gamma_{\textrm{\scriptsize{int}}}=e^{-b_t}
\left[\sum_{n=0}^{\infty}\frac{(-1)^{n}}{n!}
\frac{(b_t)^n}{\langle\Gamma_{n}\rangle}\right]^{-1},
  \end{eqnarray}
where the width $\langle\Gamma_n\rangle=\langle\Gamma_{\textrm{\scriptsize{eff}}}\rangle +2n\hbar\beta_t$ and the averaged effective width
is determined by 
\begin{equation}\label{eq:G5}
\langle\Gamma_{\textrm{\scriptsize{eff}}}\rangle =\Gamma_0+2\hbar\kappa^2D_t[1+0.3(a_c/a_h)^2]+\Delta\Gamma_N.
\end{equation}

 Significance of Langevin's approach is illustrated by Fig.2, where  the Brownian broadening $\langle\Delta\Gamma_B\rangle$ 
  for magnetite MNPs with diameter 700 nm is calculated by Eq.~(\ref{eq:22}) (curve 1) and by  Eq.~(\ref{eq:101}) (curve 2). 
  In the last case we put $b_r=0$, that is describe the translational Brownian motion by Langevin's equation and the rotational one by simple diffusion equation.
Note that in previous articles \cite{K,B2} the Brownian rotation has not been taken into consideration at all.

  \section{Discussion}
It can be easily shown that the considered relaxations do not influence on the overall probability of the absorption without recoil.
Really, since
\begin{equation}\label{}
\int_{-\infty}^{\infty}\exp(ist/\hbar)ds=2\pi\hbar\delta(t)
\end{equation}
and $G_{B}({\boldsymbol \kappa},0)G_N(0)=1$
one gets
\begin{equation}\label{}
\int_{-\infty}^{\infty}\sigma_a(s)ds=\frac{\pi}{2}\sigma_0\Gamma_ae^{-2W}.
\end{equation}
Thus, the square of the M\"{o}ssbauer spectrum  keeps constant value.

\begin{figure}[b]
\label{fig3}\vspace{-1cm}
\centerline{\includegraphics[height= 6cm, width= 9cm]{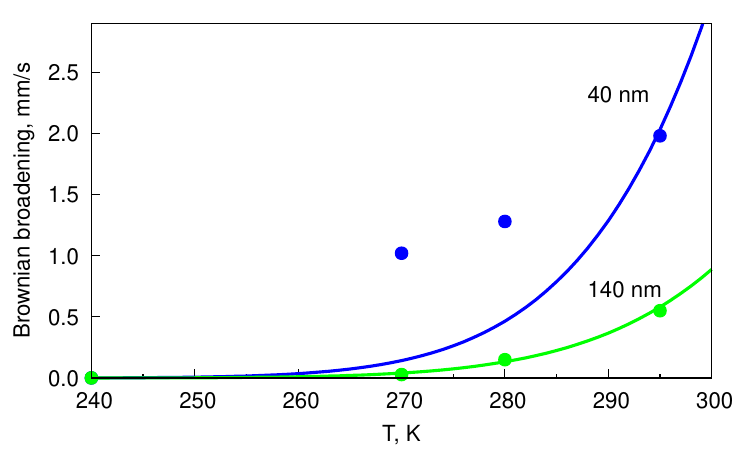}}
\vspace{-0.3cm}
\caption{ Brownian broadening of the  absorption lines vs temperature for the magnetite nanoparticles with diameters 40 nm and 140 nm dissolved in  96\%  glycerol-water mixture. The experiment \cite{Chuev1} is presented by circles, our calculations by solid curves. }
\end{figure}

We have derived the expression (\ref{eq:s1}) for   the absorption cross section $\sigma_a(s)$ of M\"{o}ssbauer radiation by MNPs suspended in a liquid. It will coincide with the result of Blume and Tion \cite{Blume}, if in the effective width $\Gamma_{\textrm{\scriptsize{eff}}}$ we omit a contribution $\Delta\Gamma_B$ of the Brownian motion. But the cross section, averaged over uniform distribution of the nuclei $^{57}$Fe over the particle, takes a combersome form. The situation significantly simplifies if we replace averaging of $\sigma_a(s)$ by the averaging only of $\Delta\Gamma_B$. As it is seen in Fig.1, this procedure provides very good result.

We compared our calculations with the experimental data \cite{Cher}  of Cherepanov {\it et al.} (see Fig.3}).  From the M\"{o}ssbauer spectra of ferrofluids   they subtracted the spectra of dried samples. It enabled them to extract  the contribution only of Brownian motion into the broadening of the spectral lines.  The calculated dependence of the Brownian broadening $\Delta\Gamma_B$ on the temperature is presented in Fig.3 by solid lines and the experimental data by circles. The calculations very well agree with the experiment  for large particles having diameter 140 nm, and at the same time they terribly  deviate for small ones with the dimensions 40 nm. This deviation contradicts to the fact that the viscosity of liquid exponentially decreases with growing temperature, and as a consequence this should ensure for small MNPs  the same behavior  of  $\Delta\Gamma_B$  as for large MNPs.

The next part of the paper will be addressed to M\"{o}ssbauer spectra in ferrofluids subjected to external magnetic fields.

\end{document}